\documentstyle[12pt]{article}

\begin{document}

\title{Computers with closed timelike curves can solve hard problems}

\author{Todd A. Brun \\
Institute for Advanced Study, \\
Einstein Drive, Princeton, NJ  08540 }

\maketitle

\begin{abstract}
A computer which has access to a closed timelike curve, and can
thereby send the results of calculations into its own past, can
exploit this to solve difficult computational problems efficiently.
I give a specific demonstration of this for the problem of factoring
large numbers, and argue that a similar approach can solve NP-complete
and PSPACE-complete problems.   I discuss the potential impact of
quantum effects on this result.
\end{abstract}

\section{Computing with closed timelike curves}

The recent success in the field of quantum computation shows how
the power of computation can be affected by the particular choice
of physical model for a computer.  By assuming a computer which
operates according to the laws of quantum mechanics, Peter Shor was
able to devise an algorithm to factor large numbers exponentially
more efficiently than the best known classical algorithm \cite{Shor}.

This success leads one to ask:  are there other physical models for
computation which will also result in much more powerful algorithms?
As long as one is speculating, one might as well speculate wildly;
so let us consider computers with access to closed timelike curves (CTCs),
which are thereby able to send information (such as the results of
calculations) into their own past light cones \cite{Morris,Friedman}.

I argue that such computers would be able to solve computationally
difficult problems with amazing (indeed, almost magical) efficiency.
In honor of Shor, I consider an algorithm for factoring large numbers;
but it is easy to see that a very large class of computationally
difficult problems, including NP-complete and PSPACE-complete problems,
can be solved by the same trick.

For purposes of illustration, I assume that this computer has a
particular register, {\tt timeRegister}, which can be set from the
future by messages sent via the CTC.  The command used for setting
{\tt timeRegister} is {\tt timeSet(t,x)}, where {\tt t} is the
time at which to set the register and {\tt x} is the value to which
it should be set.  In between resetting events, the register's
value remains unchanged.

Now consider the following program:
\begin{verbatim}
        input(N);
        timeRegister = -1;
        t = clock();
        p = timeRegister;
        if (p > 1) and (N mod p = 0) go to FINAL;
        p = 1;
        do
          p = p + 1;
        until (N mod p = 0) or (p > sqrt(N));
        if (p > sqrt(N)) p = N;
 FINAL  timeSet(t,p);
        output(p);
        end;
\end{verbatim}
At the start of the program, the computer checks the value of
{\tt timeRegister}; if it divides {\tt N}, the computer skips
to the end of the program, sends the answer back in time, and outputs it.
Otherwise the computer exhaustively searches until it finds a factor of
{\tt N}, then sends the factor back in time and outputs it.

What happens when this program is executed?  If {\tt timeRegister} does
{\it not} divide {\tt N} when it is checked, the computer
searches until it finds a factor, then sends it back; so
{\tt timeRegister} {\it will} divide {\tt N} when it is checked,
which is a contradiction.  On the other hand, if {\tt timeRegister}
{\it does} divide {\tt N} when it is checked, it will skip
to the end, send the factor back, and quit.  While this situation is
quite bizarre, it is self-consistent; and if only {\it self-consistent}
evolutions can occur, then this must be what happens
\cite{Novikov,Earman95a,Earman95b,Carlini95,Konstantinov,Carlini96,Romero}.

The operation of this computer is reminiscent of an old time-travel
paradox.  The brilliant young inventor receives a message from her future
self, telling her that she is going to invent a time machine, and giving
her the details of its construction.  She duly builds the machine and
demonstrates it.  When she is old and famous, she sends a message back
to her younger self, telling her that she is going to invent a time
machine, and giving her the details of its construction.

This situation is self-consistent, but still very strange; the information
on how to build a time machine appears out of nowhere.  On the other hand,
if the information {\it hadn't} appeared, that would have been self-consistent
as well; the young inventor need not have ever discovered the time machine.

In the computer program, by contrast, there is a contradiction if the
information {\it doesn't} appear.  This situation is created by the inner
search loop, which is guaranteed to find the answer sooner or later.
Therefore this inner loop is necessary for the algorithm to function,
even though the loop itself will never be executed.

\section{A recursive version}

There are a couple of loopholes in this argument that need to be
addressed.  The contradiction depends on the ability of the computer
to execute the inner loop and find an answer.  It is not very difficult,
however, to choose {\tt N} so large that it would take longer than the
lifetime of the universe to find a factor by exhaustive search.  It is
quite easy to choose {\tt N} big enough that it will take longer than
the lifetime of any reasonable computer.  If the program never gets to
the final {\tt timeSet} command, it will never send the factor back,
and no contradiction can arise.

A similar problem might arise if the CTC doesn't extend arbitrarily far
into the future.  A CTC with limited extent might only be able to send
information back a short distance in time; if the inner loop only
finishes executing in the future of the CTC, the program will be unable
to send the factor back, and once again no contradiction arises.

This might seem an insuperable problem, but I believe that it can be
circumvented by using a more clever algorithm than exhaustive search.
Consider the following function:

\begin{verbatim}
        function factor(N,nStart,nEnd)
          timeRegister=-1;
          t = clock();
          p = timeRegister;
          if (p=0) or (N mod p = 0) go to FINAL;
          if (nEnd-nStart < nTractable)
            p = nStart-1;
            do
              p = p + 1;
            until (N mod p = 0) or (p > nEnd);
            if (p > nEnd) p = 0;
          else
            p = factor(nStart,(nStart+nEnd)/2);
            if (p = 0) 
              p = factor(1+(nStart+nEnd)/2,nEnd);
 FINAL    timeSet(t,p);
          return p;
        end function;
\end{verbatim}
The function {\tt factor(N,nStart,nEnd)} looks for a factor of
{\tt N} within the range {\tt nStart} to {\tt nEnd}.  If there
is one, then it returns the factor; if not, it returns 0.

The function first checks to see if an answer has been sent back from
the future.  If one has, it skips to the end and returns it.  If not,
it checks if the range {\tt nEnd-nStart} is small enough to search
in a small time (determined by a constant parameter {\tt nTractable}).
If the range is small enough, it loops until it finds an answer.
If not, it breaks the range into two parts and calls itself
recursively for each subrange.  At the end, it sends the answer back in
time and returns it.

The factoring program then takes the following form:

\begin{verbatim}
        input(N);
        timeRegister = -1;
        t = clock();
        p = timeRegister;
        if (N mod p = 0) go to FINAL;
        p = factor(N,2,sqrt(N));
        if (p = 0) p = N;
 FINAL  timeSet(t,p);
        output(p);
        end;
\end{verbatim}
Once again, at the beginning the program checks to see if the answer
has been sent back from the future.  If it has, then it skips to the
end, sends the answer back in time, and outputs it.  If not, then
it enters the recursion.  At the bottom level of the recursion is a loop
that can be executed in a short time.  The loop is only executed if
the result of the loop is {\it not} sent back in time, but if the loop
is executed then the result {\it will} be sent back in time.  Therefore
the loop will {\it not} be executed, and the answer will appear when
checked.  At the next higher level of recursion, the call
to {\tt factor} {\it won't} be made, because the answer there will already
have appeared; and so forth, all the way to the top of the recursion.
The function {\tt factor} will actually never be called at all.
The only self-consistent outcome is that the program finds
the correct answer when it checks {\tt timeRegister} at the beginning.

The only requirement for this program to work is that the number of
recursive calls to break the interval down into a tractable subinterval
not be too big.  Since the number of levels of recursion goes like
$\log_2 {\tt N}$, this is not very restrictive.

\section{Harder problems}

The particular algorithm I presented solved the factoring problem.
While this problem is in NP, it is not NP-complete; but it is obvious
that a program with the same structure could solve NP-complete problems
as well.  Indeed, it can solve even more difficult problems, as we shall
see.

First, consider the {\it satisfiability} problem (SAT), which is to find
a string of {\tt N} bits $x_1,\ldots,x_N$ which simultaneously make
true (satisfy) some set of {\it clauses} (i.e., logical statements) $\phi$.
These clauses $\phi$ can be put together into a single logical statement
involving the $\{x_i\}$, in conjunctive normal form.  For example,
\begin{equation}
\phi = (x_1 \vee \neg x_3 \vee x_{41}) \wedge (\neg x_5 \vee x_{17}) \wedge
  \cdots \;.
\end{equation}

The satisfiability
problem can be solved (very inefficiently) by exhaustive search, merely
trying every {\tt N}-bit string until either finding one that satisfies
the clauses, or determining that there isn't one.  By breaking down the
set of all strings into smaller and smaller subsets using a recursive
algorithm, one could modify the program in section II to solve SAT.

SAT is the canonical example of an {\it NP-complete} problem
\cite{Papadimitriou}.  Any problem in NP can be translated into an
instance of the satisfiability problem with only polynomial overhead.
Therefore, SAT is in that sense at least as difficult as any other
problem in NP.  However, these are not necessarily the most difficult
problems that exist.  Consider the following variant of the problem.

Once again we have a set of {\tt N} bits and a set of clauses $\phi$.
This time, however, we don't want to know if there is an assignment of
bit values that satisfies $\phi$;  instead, we want to know if there
is a value of $x_1$ such that for all $x_2$ there is a value of $x_3$
such that for all $x_4$, etc., such that $phi$ is satisfied:
\begin{equation}
\exists x_1 \forall x_2 \exists x_3 \forall x_4 \cdots
  \exists x_{N-1} \forall x_N \phi \;,
\end{equation}
where I have assumed that {\tt N} is even.

This is the {\it quantified satisfiability} problem (QSAT) \cite{QSAT}.
There is no known NP algorithm to solve QSAT; it is an example of a
{\it PSPACE-complete} problem, i.e., a problem which can be solved using
a computer with an amount of space polynomial in {\tt N}, and which is
polynomially equivalent to all other such problems.  PSPACE-complete
problems are believed to be strictly harder than NP-complete problems.

Such problems can be solved recursively.  Suppose that ${\tt QSAT}(\phi)$
is the function that evaluates QSAT for the set of clauses $\phi$.
Define $\phi_{00},\phi_{01},\phi_{10},\phi_{11}$, where $\phi_{ij}$ is
the set of clauses $\phi$ with $x_1x_2$ replaced by $ij$.  Then we see
that
\begin{equation}
{\tt QSAT}(\phi) = ( {\tt QSAT}(\phi_{00}) \wedge {\tt QSAT}(\phi_{01}) )
  \vee ( {\tt QSAT}(\phi_{10}) \wedge {\tt QSAT}(\phi_{11}) ) \;.
\label{QSAT_recursion}
\end{equation}
The instances of QSAT on the right-hand side of (\ref{QSAT_recursion})
are all of length $N-2$, and can be replaced by similar recursive
expressions.  By making use of this recursive structure, we can modify
the program in section II to solve QSAT.  Therefore, computers with
CTCs should be able to solve not only NP-complete, but even
PSPACE-complete problems efficiently.

One interesting difference in this case, however, is that while it
is simple to check the answer to SAT (just by checking that the returned
set of bit values does indeed satisfy $\phi$), there is no efficient
way of checking that an answer to QSAT is correct, in general.

\section{Quantum considerations}

So far in this paper I have treated both the computer and the CTC
as if they were completely classical.  I have adopted the assumption
\cite{Novikov,Carlini95} that the allowed evolutions are those which
do not produce a contradiction, and that the computer functions
deterministically.  How will these results change when we take
quantum mechanics into account?

Of course, the most likely outcome of including quantum effects is
that CTCs will no longer exist at all.  This is the so-called
``Chronology Protection Conjecture'' of Steven Hawking, which postulates
that the build-up of quantum fluctuations around a CTC will destabilize
the spacetime and destroy the time machine \cite{Hawking}.  There is some
evidence to support this conjecture, though there is (as yet) no proof.

Several authors have investigated quantum systems on a fixed background
spacetime which includes CTCs.  (See, e.g., \cite{Politzer}
and \cite{Rosenberg}.)  David Deutsch \cite{Deutsch},
in particular, has suggested that the Many-Worlds
interpretation of quantum mechanics prevents time travel paradoxes.
When one travels back in time to kill one's grandfather (in the usual
violent version of the paradox), one finds oneself in a different branch
of the wavefunction; the future of the new ``world'' is changed, but not
the old ``world.''

Would this kind of argument eliminate a computer such as I describe?  It
is not obvious that it would.  Deutsch's argument prevents contradictions,
but the operation of the algorithm, while mind-boggling, is not
contradictory.  The operation of the computer is deterministic, and
should proceed identically in (almost) all universes in which it occurs
(barring very improbable quantum fluctuations which, for example, demolish
the lab).  There seems no reason that self-consistent worlds with
causal loops cannot exist.  They don't defy logic, but only common sense.
Indeed, Deutsch himself in \cite{Deutsch} suggested that closed timelike
curves might make possible computers which solve hard problems.

It is possible that some other quantum effect might prevent the algorithm
from working, while still allowing the existence of CTCs.  But at present,
no such argument has occurred to me.

\section{Conclusions}

It is very odd for information to suddenly appear out of nowhere, but
in a universe with closed timelike curves such events can be
expected to occur.  It has widely argued that if CTCs are possible,
the laws of physics should require that only internally consistent
evolutions can occur, and that generalized versions of the principle
of least action will enforce this behavior
\cite{Novikov,Carlini95,Konstantinov,Carlini96}.

I've argued in this paper that one could exploit
this tendency to design computers able to solve hard problems in
very little time.  I gave the specific example of factoring; but
in section III, I argued that a similar algorithm could solve both
NP-complete and PSPACE-complete problems as well, using the satisfiability
and quantified satisfiability problems as examples.  In all these cases,
the answers appear out of nowhere in order to prevent logical
contradictions from arising.  Thus, these algorithms can be said to
work because of the presence of brute-force search loops which are never
actually executed.

This is a strange, though logically consistent, conclusion.  But perhaps
the best conclusion to draw is that it makes the existence
of closed timelike curves even more unlikely.

\section*{Acknowledgments}
I would like to thank Steve Adler, Avi Wigderson, and Andrew Yao for
reading drafts of this manuscripts and giving very helpful feedback;
and Cara King, for encouraging me to write up this idea is spite of
its innate craziness.  This work was supported in part by the
Martin A.~and Helen Chooljian Membership in Natural Sciences, and by
DOE Grant No.~DE-FG02-90ER40542.

\end{document}